\documentclass[twocolumn,showpacs,showkeys,nofootinbib,aps,prd,letterpaper,floatfix,superscriptaddress]{revtex4-1}

\usepackage{amsmath}
\usepackage{amssymb}
\usepackage{multirow}
\usepackage{graphicx}
\usepackage{float}
\usepackage{appendix}
\usepackage{color}
\usepackage{url}
\usepackage{subfigure}
\usepackage{slashed}
\usepackage{enumerate}

\def\nl{\nonumber\\}
\def\cal{\mathcal}

\renewcommand{\Im}{\textrm{Im}\,}
\newcommand{\GeV}{{\,\textrm{GeV}}}
\newcommand{\Order}{\mathcal{O}}

\RequirePackage{xspace} \allowdisplaybreaks

\begin{document}

\title{\boldmath $B_{l4}$ decays and the extraction of $|V_{ub}|$}

\author{Xian-Wei Kang}\email{x.kang@fz-juelich.de}
\affiliation{Institute for Advanced Simulation and
             J\"ulich Center for Hadron Physics,
             Institut f\"ur Kernphysik,
             Forschungszentrum J\"ulich,
             52425 J\"ulich, Germany}

\author{Bastian Kubis}\email{kubis@hiskp.uni-bonn.de}
\affiliation{Helmholtz-Institut f\"ur Strahlen- und Kernphysik (Theorie) and\\
             Bethe Center for Theoretical Physics,
             Universit\"at Bonn,
             53115 Bonn, Germany}

\author{Christoph Hanhart}\email{c.hanhart@fz-juelich.de}
\affiliation{Institute for Advanced Simulation and
             J\"ulich Center for Hadron Physics,
             Institut f\"ur Kernphysik,
             Forschungszentrum J\"ulich,
             52425 J\"ulich, Germany}

\author{Ulf-G.\ Mei\ss ner}\email{meissner@hiskp.uni-bonn.de}
\affiliation{Institute for Advanced Simulation and
             J\"ulich Center for Hadron Physics,
             Institut f\"ur Kernphysik,
             Forschungszentrum J\"ulich,
             52425 J\"ulich, Germany}
\affiliation{Helmholtz-Institut f\"ur Strahlen- und Kernphysik (Theorie) and\\
             Bethe Center for Theoretical Physics,
             Universit\"at Bonn,
             53115 Bonn, Germany}

\begin{abstract}
The Cabibbo--Kobayashi--Maskawa matrix element $|V_{ub}|$ is not well determined yet. 
It can be extracted from both inclusive or exclusive decays, like $B\to\pi(\rho)l\bar\nu_l$.
However, the exclusive determination from $B\to\rho l\bar\nu_l$, in particular, suffers from a large model dependence.
In this paper, we propose to extract $|V_{ub}|$ from the four-body semileptonic decay
$B\to\pi\pi l \bar\nu_l$, where the form factors for the pion--pion system are treated in
dispersion theory. This is  a model-independent approach that takes
into account the $\pi\pi$ rescattering effects, 
as well as the effect of the $\rho$ meson. We demonstrate that
both finite-width effects of the $\rho$ meson as well as scalar
$\pi\pi$ contributions can be considered completely in this way.
\end{abstract}



\maketitle

\section{Introduction}
Precisely determining the elements of the Cabibbo--Kobayashi--Maskawa (CKM) matrix~\cite{CKM}
plays a very important role in testing the Standard Model.
Any deviations from the unitarity of the CKM matrix would be viewed as a sign
of new physics. 
The element $|V_{ub}|$ has been measured from
inclusive charmless semileptonic $B$ decay as well as
from the exclusive decays $B\to \pi(\rho) l\bar\nu_l$.
For a review on the determination of $|V_{ub}|$, see Ref.~\cite{pdg}. The
value of $|V_{ub}|$ preferred by the current global analysis of CKM
data is about 15\% smaller than the one from inclusive charmless
semileptonic $B$ decays~\cite{HFAG,Bona,CKMfitter}, a problem unresolved to date.
Furthermore, the inclusive determinations of $|V_{ub}|$ are about
two standard deviations larger than those obtained from
$B\to\pi l\bar\nu$, presently with a smaller uncertainty.
The value of $|V_{ub}|$ predicted from the measured CKM angle $\sin2\beta$, however,
is closer to the exclusive result~\cite{Bonajhep},
and it should be stressed that various theoretical extractions 
based on exclusive decays are remarkably consistent among each other~\cite{HFAG,CLEO1,CLEO2,BaBar2003,BaBar2011}.
These discrepancies prompted a reexamination of the sources of theoretical
uncertainty in the inclusive determination~\cite{discrepancy,flavor}.

In the present paper, we investigate the four-body
semileptonic decay mode $B^-\to \pi^+\pi^- l^-\bar\nu_l$ (which we will abbreviate as $B_{l4}$ for short)
and propose a method that allows one to extract $|V_{ub}|$ in a model-independent way.
As a major step forward to a reliable treatment of the hadron-physics aspects of this decay,
we use an approach based on dispersion theory
without the need to explicitly match on specific resonance contributions or to separate these from nonresonant background.
This presents a significant improvement compared to
previous studies of $B\to \rho l\bar\nu_l$~\cite{BtorhoTh}, and should serve as a valuable cross-check for
the inclusive determination.
In the future the distributions derived below could be used directly in the Monte Carlo generators of the experiments.

We include the kinematic range for invariant masses of the $\pi\pi$ pair below the $K\bar K$ threshold in our analysis
and expand the form factors for the full $B_{l4}$ transition matrix element in $\pi\pi$ partial waves up to $P$~waves; 
$D$ and higher partial waves have been checked to be negligible at these energies. 
While this model-independent description of the form factor dependence on the $\pi\pi$ invariant mass is 
in principle general and holds for arbitrary dilepton invariant masses, in practice we make use
of matching to heavy-meson chiral perturbation theory to fix the normalization of the matrix element---a prerequisite 
for the extraction of $|V_{ub}|$. 
This scheme applies in the kinematics where heavy-quark
effective field theory is valid, i.e.\ for very large dilepton invariant masses.
We point to Ref.~\cite{Siegen} for a lucid illustration of the different effective theories applicable in 
different kinematic regimes for this decay.

This manuscript is organized as follows. In Sec.~\ref{kinematics} the
kinematics for the process of the four-body semileptonic $B$ decay is
reviewed, and the form factors for the hadronic transition of
$B\to\pi\pi l\bar\nu_l$ are defined. In Sec.~\ref{ffdr}, we show in detail how to treat these form
factors within dispersion theory:
the analytic properties are summarized in Sec.~\ref{sec:analytic} and
the required pole terms calculated in heavy-meson chiral perturbation theory in Sec.~\ref{secpole},
before we provide the expressions for the various form factors in the Omn\`es representation
in Sec.~\ref{Omnes}. 
We discuss the required matching to leading-order heavy-meson chiral perturbation theory in Sec.~\ref{sec:matching}.
Numerical results are discussed in Sec.~\ref{sec:results}, and
we summarize our findings in Sec.~\ref{sec:summary}.
Some technical details are relegated to the Appendices.

\section{Kinematics, form factors, partial waves, decay rates}\label{kinematics}

\begin{figure}
\centering
\includegraphics[width=\linewidth]{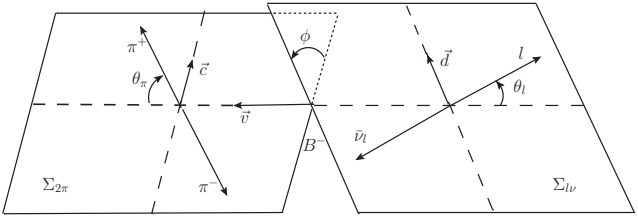}
\caption{Illustration of the kinematical variables for $B_{l4}$.} \label{decayplane}
\end{figure}
The kinematics of the process $B^-(p_B)\to \pi^+(p_+)\pi^-(p_-) l^-(p_l) \bar \nu_l(p_\nu)$ 
are described in terms of the five variables displayed 
in Fig.~\ref{decayplane}~\cite{Bijnens,Cabibbo,Berends}:
\begin{enumerate}[(i)]
\item the effective mass squared of the pion pair $s=(p_++p_-)^2 = M_{\pi\pi}^2$;
\item the effective mass squared of the dilepton pair $s_l = (p_l+p_\nu)^2$;
\item the angle $\theta_\pi$ of the $\pi^+$ in the $\pi^+\pi^-$ center-of-mass frame
$\Sigma_{2\pi}$ with respect to the dipion line of flight in the $B^-$ rest frame $\Sigma_B$;
\item the angle $\theta_l$ of the charged lepton $l$ in the lepton center-of-mass system $\Sigma_{l\nu}$ with
respect to the dilepton line of flight in $\Sigma_B$;
\item the angle $\phi$ between the dipion and dilepton planes.
\end{enumerate}
Two additional Mandelstam variables are defined as
\begin{align}
t&=(p_B-p_+)^2\,,\quad u=(p_B-p_-)^2\,,\nl
\Sigma_0&\equiv s+t+u=2M_\pi^2+m_B^2+s_l\,. \label{Sigma0}
\end{align}
We define the combinations of four vectors $P=p_++p_-$, $Q=p_+-p_-$, $L=p_l+p_\nu$,
and make use of the kinematical relations
\begin{equation}\label{costhetapi}
(PL)\equiv P\cdot L=\frac{m_B^2-s-s_l}{2}\,,\quad t-u=-2\sigma_\pi X\cos\theta_\pi \,,
\end{equation}
where
\begin{equation}\label{definitionX}
\sigma_\pi=\sqrt{1-\frac{4M_\pi^2}{s}}\,,\quad
X=\frac{1}{2}\lambda^{1/2}(m_B^2,s,s_l) \,,
\end{equation}
and the K\"all\'en triangle function is given by
$
\lambda(a,b,c)=a^2+b^2+c^2-2(ab+ac+bc)
$.

We decompose the matrix element in terms of form factors according to
\begin{align}
T&=\frac{G_F}{\sqrt{2}}V^*_{ub}\bar v(p_\nu)\gamma^\mu(1-\gamma_5)u(p_l)I_\mu \,, \nl
I_\mu&= \langle\pi^+(p_+)\pi^-(p_-)|\bar u\gamma_\mu(1-\gamma_5)b|B^-(p_B)\rangle \label{eq:FGHRdef}\\
&= -\frac{i}{m_B}(P_\mu F+Q_\mu G+L_\mu R)-\frac{H}{m_B^3}\epsilon_{\mu\nu\rho\sigma}L^\nu P^\rho Q^\sigma \,, \nonumber
\end{align}
where $G_F=1.166365\times 10^{-5}$\,GeV$^{-2}$  is the Fermi
constant, and we use the convention $\epsilon_{0123}=1$. The first three terms correspond to the axial
current part, whereas the last term corresponds to the vector current.
The dimensionless form factors $F$, $G$, $H$, and $R$ are analytic functions
of three independent variables, e.g.\ $s$, $s_l$, and $t-u$. 
Their partial-wave expansions for fixed $s_l$ read~\cite{Bijnens,Berends}
\begin{align}\label{FGRHPWE}
F&=\sum_{l\geq 0} P_l(\cos\theta_\pi) f_l-\frac{\sigma_\pi(PL)}{X}\cos\theta_\pi G\,,\nl
G&=\sum_{l\geq 1} P'_l(\cos\theta_\pi)g_l\,,\quad
H=\sum_{l\geq 1} P'_l(\cos\theta_\pi)h_l\,, \nl
R&=\sum_{l\geq 0} P_l(\cos\theta_\pi)r_l+\frac{\sigma_\pi s}{X}\cos\theta_\pi G\,, 
\end{align}
where $P_l(z)$ are the standard Legendre polynomials and
$
P'_l(z)={d}P_l(z)/dz
$.
An alternative set of form factors is given by
\begin{align}\label{orthff}
F_1&=X\cdot F+\sigma_\pi (PL)\cos\theta_\pi G \,, \quad
F_2=G\,, \quad F_3=H\,, \nl 
F_4&=-(PL)F-s_l R-\sigma_\pi X\cos\theta_\pi G \,,
\end{align}
whose partial-wave expansions 
\begin{align}
F_1&=X \sum_{l\geq 0} P_l(\cos\theta_\pi) f_l\,, &
F_2 &=\sum_{l\geq 1}P'_l(\cos\theta_\pi) g_l \,,\nl
F_3&=\sum_{l\geq 1}P'_l(\cos\theta_\pi)h_l \,, &
F_4&=\sum_{l\geq 0}P_l(\cos\theta_\pi)\tilde{r}_l \,,\nl
&& \tilde{r}_l&=-\big((PL)f_l+s_lr_l\big) \,,
\label{eq:pw-expansion}
\end{align} 
directly follow from Eqs.~\eqref{FGRHPWE} and \eqref{orthff}. 
Note that all partial waves $f_l$, $g_l$, $h_l$, $r_l$ ($\tilde r_l$) 
are functions of $s$ and $s_l$.
The lowest angular-momentum
$\pi\pi$ state contributing to the form factors $F_2$ and $F_3$ is
the $P$-wave state, whereas the form factors $F_1$ and $F_4$ start
with $S$~waves. For the partial-wave decomposition up to $P$~waves,
we can therefore write
\begin{align}\label{PWD}
F_1&=X\big[f_0(s,s_l)+f_1(s,s_l)\cos\theta_\pi+\ldots \big] \,,\nl
F_2&=g_1(s,s_l)+\ldots\,,\qquad F_3=h_1(s,s_l)+\ldots\,,\nl
F_4&=\tilde r_0(s,s_l)+\tilde r_1(s,s_l)\cos\theta_\pi+\ldots\,,
\end{align}
where the ellipses denote higher partial waves. In the following, we
sometimes suppress the dependence on $s_l$ in order to ease
notation.

The decay rate, after integration over the angles $\phi$ and
$\theta_l$, reads
\begin{align}\label{dgamma3}
d\Gamma&=G_F^2|V_{ub}|^2N(s,s_l)J_3(s,s_l,\theta_\pi)ds\,ds_l\,d\cos\theta_\pi
\,,\nl J_3(s,&s_l,\theta_\pi)=\frac{2+z_l}{3}|F_1|^2 +z_l|F_4|^2 \nl
&+\frac{(2+z_l)\sigma_\pi^2s\,s_l}{3}\bigg(|F_2|^2+\frac{X^2}{m_B^4}|F_3|^2\bigg)\sin^2\theta_\pi
\,,
\end{align}
with
\begin{equation}
z_l=\frac{m_l^2}{s_l}\,,\quad
N(s,s_l)=\frac{(1-z_l)^2\sigma_\pi X}{2(4\pi)^5m_B^5}\,.
\end{equation}
In most of the available phase space (including the kinematic
regime where chiral perturbation theory can be applied), the mass of
the lepton can be neglected (i.e.\ $z_l\ll 1$), and the
contribution of $F_4$ to the decay rate is therefore invisible in particular
for  $B_{e4}$ decays, since it is always associated with a factor of $z_l$.
We will not analyze the form factor $F_4$ and its partial waves $\tilde r_i$ in the following.
Integrating Eq.~\eqref{dgamma3} over $\cos\theta_\pi$ yields
the partial decay rate $d\Gamma/(ds\,ds_l)$; neglecting terms of order $z_l$
and inserting the partial-wave expansions Eq.~\eqref{eq:pw-expansion}, we find
\begin{align}\label{partialwidth}
\frac{d\Gamma}{ds\,ds_l}&=G_F^2|V_{ub}|^2N(s,s_l)J_2(s,s_l)\,,\nl
J_2(s,s_l)&=\int_{-1}^1 d\cos\theta_\pi J_3(s,s_l,\cos\theta_\pi)
\nl &= \frac{4X^2}{3}\bigg(|f_0(s)|^2+\frac{1}{3}|f_1(s)|^2\bigg)\nl
& +\frac{8}{9}\sigma_\pi^2
s\,s_l\bigg(|g_1(s)|^2+\frac{X^2}{m_B^4}|h_1(s)|^2\bigg) +\ldots \,,
\end{align}
where the ellipsis denotes the neglected $D$ and higher waves.
Interference terms between different partial waves vanish
upon angular integration, such that the partial-wave contributions to the decay rate can be easily read off.

\section{Form factors in dispersion theory}\label{ffdr}

\subsection{Analytic properties}\label{sec:analytic}

The principle of maximal analyticity, which states that amplitudes
possess no other singularities than those stemming from unitarity
and crossing~\cite{Queen}, tells us that the partial-wave amplitudes
$f_l$, $g_l$, and $h_l$ have the following analytic properties.
\begin{enumerate}[(i)]
\item At fixed $s_l$, they are analytic in the complex $s$ plane, cut
along the real axis for $s\geq 4M_\pi^2$ and $s\leq 0$. 
The presence of left-hand cuts $s\leq0$ follows from the relations 
\begin{align}
& t=\frac{\Sigma_0-s}{2}-\sigma_\pi X\cos\theta_\pi\,,\nl &
t(\cos\theta_\pi=-1, s < 0)\geq (m_B+M_\pi)^2\,
\end{align}
(and equivalent expressions for $u$),
since the form factors $F$, $G$, and $H$ have cuts for
$t,\,u\geq(m_B+M_\pi)^2$.
\item In the interval $0\leq s\leq 4M_\pi^2$, they are real.
\item In the interval $4M_\pi^2\leq s\leq 16M_\pi^2$, Watson's theorem~\cite{Watson}
is satisfied and therefore the phases of the partial-wave amplitudes
($f_l$, $g_l$, $h_l$) coincide with the corresponding pion--pion
scattering phases.
\item For the crossed ($t$ and $u$)
channels, due to the lack of experimental information on $\pi B$
phase shifts, we will approximate the $\pi B$ interaction by $B^*$
pole terms.
\end{enumerate}
In practice, the range of validity of Watson's theorem can be extended to
a larger domain, e.g.\ for the $S$~wave to $s\leq s_K=4M_K^2\approx 1$\,GeV$^2$,
since inelasticities due to four or more pions are strongly suppressed both by phase space
and by chiral symmetry. As pointed out e.g.\ in Refs.~\cite{Donoghue_scalarff,GasserMeissner},
chiral perturbation theory predicts the inelasticity parameter of the $\pi\pi$ $S$ and $P$~waves
to be of order $p^8$ below the $K\bar K$ threshold,
while the corresponding scattering phase shifts are of order $p^2$.
Phenomenological analyses of the $\pi\pi$
interactions show that final states containing more than two
particles start playing a significant role only well above the
$K\bar K$ threshold $s_K$~\cite{simon}. Here we refrain from performing
a coupled-channel study, which limits the applicability of our approach to the region below $s_K$.
The subtleties associated with the strong onset of inelasticities in the $S$~wave
in the vicinity of $s_K$ (very close to the $f_0(980)$ resonance)
for scalar form factors of the pion will be briefly discussed in
Sec.~\ref{sec:phases}.

\subsection{Heavy-meson chiral perturbation theory}\label{secpole}

\begin{figure}
\centering
\includegraphics[scale=0.9]{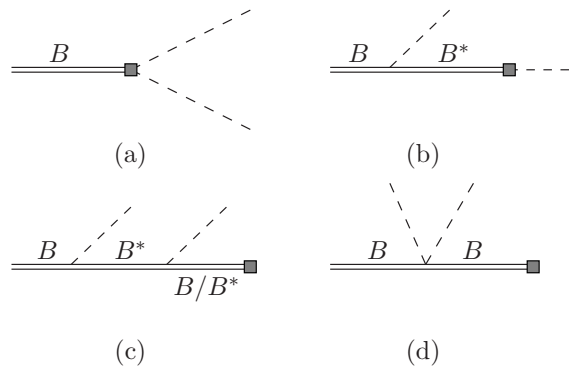}
\caption{Leading-order diagrams for $B\to\pi\pi$ matrix elements of
the hadronic current. Diagrams (b) and (c) contain $u$-channel pole
terms. Solid double lines and dashed lines represent heavy mesons and
pseudo-Goldstone bosons, respectively. The shaded square denotes
an insertion of the left-handed leptonic current. Diagram (c) involves
both $BB^*\pi$ and $B^*B^*\pi$ vertices.
Diagrams (a) and (d) are suppressed in the chiral expansion as long as
the lepton mass is neglected.}\label{fig:treediagram}
\end{figure}

In the process $B^-\to \pi^+\pi^- l^-\bar\nu_l$,
$u$-channel contributions contain pole terms, while
$t$-channel contributions do not. We obtain the pole
terms by computing the leading-order diagrams (b) and (c) of Fig.~\ref{fig:treediagram} 
in the framework of heavy-meson chiral perturbation theory~\cite{Wise,Dl4,HMChPTPR}.
Let us briefly review the heavy-meson chiral Lagrangian. Define the
heavy-meson field and its conjugate as
\begin{equation}
H_a = \frac{1+\slashed v}{2}\big(P_{a\mu}^*\gamma^\mu-P_a\gamma_5\big) \,, \quad
\bar H_a = \gamma^0 H^\dagger_a\gamma^0\,,
\end{equation}
where $P_{a\mu}^*$ is the field operator that annihilates a $P_a^*$ meson
with velocity $v$, satisfying $v^\mu P_{a\mu}^*=0$, and
$P_a$ annihilates a $P_a$ meson of velocity $v$. For the $B$ meson family,
we have
\begin{align}
(P_1,\,P_2,\,P_3)&=(B^-,\,\bar B^0,\,\bar B_s^0) \,,\nl
(P_1^*,\,P_2^*,\,P_3^*)&=(B^{*-},\,\bar B^{*0},\,\bar B^{*0}_s) \,,
\end{align}
which have dimension [mass]$^{3/2}$. The light pseudoscalar Goldstone boson fields are organized in
\begin{equation}
u=\exp\Big(\frac{i\phi}{2f_\pi}\Big) \,,
\end{equation}
 with
\begin{equation}
\phi=\sqrt{2}\begin{bmatrix}\frac{\pi^0}{\sqrt{2}} + \frac{\eta}{\sqrt{6}} &
\pi^+& K^+\\ \pi^- &-\frac{\pi^0}{\sqrt{2}} +
\frac{\eta}{\sqrt{6}} & K^0\\ K^-  &\bar K^0
&-\sqrt{\frac{2}{3}}\eta\end{bmatrix}.
\end{equation}
$f_\pi\simeq 92.2$\,MeV is the pion decay
constant~\cite{PDGdecayconst}. Based on these building blocks, the
leading-order Lagrangian describing the interactions of the $B$
family and the Goldstone bosons reads~\cite{Wise}
\begin{align}\label{eq:LpiB}
\mathcal{L}&=-i\,\textrm{Tr}\bar H_a v_\mu \partial^\mu H_a
+\frac{1}{2}\textrm{Tr}\bar H_aH_b
v^\mu\big(u^\dagger\partial_\mu u+u\partial_\mu u^\dagger\big)_{ba} \nl& \quad
+\frac{ig}{2} \textrm{Tr}\bar
H_aH_b\gamma_\nu\gamma_5\big(u^\dagger\partial^\nu u-u\partial^\nu u^\dagger\big)_{ba}\,.
\end{align}
Determining the coupling $g=g_{B^*B\pi}=g_{B^*B^*\pi}$, using
heavy-quark symmetry, from the partial decay width for $D^{*+}\to D^0\pi^+$ leads to
$g = g_{D^*D\pi}=0.58\pm0.07$, with the error given by the uncertainty in the width of the $D^{*+}$.
This is in surprisingly good agreement with the most recent lattice simulations, which find  
$g_{B^*B\pi}=0.516\pm 0.052$~\cite{Ohki}  and
$g_{B^*B\pi}=0.569\pm 0.076$~\cite{UKQCD} (we have added different error sources in quadrature for simplicity in both cases). 
In the present analysis, we stick to the experimental number extracted from $D^{*+}$ decays for illustration.
The dominant parts of the $B_{l4}$ amplitude will depend on $g$ in a very simple manner
(being directly proportional either to $g$ or to $g^2$), thus suggesting a straightforward strategy
towards an extraction of $|V_{ub}|$ via lattice calculations of $g_{B^*B\pi}$.

To improve on the analytic properties of the amplitudes calculated in heavy-meson chiral perturbation theory,
we include the effect of the $B^*$--$B$ mass splitting,
defined by $\Delta=m_{B^*}-m_B$ (which is of $\mathcal{O}(1/m_Q)$),
in the propagators, which in the heavy-meson approximation are of the form
\begin{align}\label{HMpropagator}
\frac{i}{2v\cdot k} & \quad \textrm{for the pseudoscalar $B$ meson},\nl
\frac{-i(g_{\mu\nu}-v_\mu v_\nu)}{2(v\cdot k-\Delta)} & \quad \textrm{for the vector $B^*$},
\end{align}
where $k$ is the small residual momentum of the propagating $B$ or $B^*$.  
We do not otherwise include heavy-quark-symmetry-breaking effects, and stick to Eq.~\eqref{eq:LpiB}
for the determination of the interaction vertices.

The left-handed current $L_{\nu a}=\bar q_a\gamma_\nu(1-\gamma_5)Q$, with $q_a$ denoting
a light and $Q$ the heavy quark, is written in chiral perturbation theory as
\begin{equation}
L_{\nu a}=\frac{i\sqrt{m_B}f_B}{2}\textrm{Tr}\big[\gamma_\nu (1-\gamma_5)H_bu^\dagger_{ba}\big]+\ldots \,, \label{eq:current}
\end{equation}
where the ellipsis denotes terms with derivatives, factors of the light-quark mass matrix $m_q$, or factors of $1/m_Q$. 
Computing the trace, one can write it explicitly as
\begin{equation}
L_{\nu a}=i\sqrt{m_B}f_B (P_{b\nu}^*-v_\nu P_b)u^\dagger_{ba}+\ldots \,.
\end{equation}
$f_B$ is the $B$ meson decay constant;
averaging the most recent lattice calculations with $2+1$ dynamical quark flavors
leads to the very precise value $f_B= 190.5\pm4.2$\,MeV~\cite{FLAG}. 
The whole $B_{l4}$ decay amplitude is proportional to $f_B$, such that any uncertainty on this parameter
directly translates into a contribution to the error in the extraction of $|V_{ub}|$.

We briefly discuss the chiral power counting of the $B_{l4}$ amplitudes and form factors.
If we denote soft pion momenta, or derivatives acting on the pion field, by $p$ generically, 
the current of Eq.~\eqref{eq:current} is $\Order(p^0)$, and so we expect to be the leading-order
amplitude resulting from the diagrams in Fig.~\ref{fig:treediagram}. 
Equation~\eqref{eq:FGHRdef} then suggests the leading contributions to the form factors $F$, $G$, $H$, and $R$
to be of chiral orders $p^{-1}$, $p^{-1}$, $p^{-2}$, and $p^0$, respectively (remember that the 
dilepton momentum $L_\mu$ is large, of order $m_B$); the alternative form factors $F_1$ and $F_4$ both
are $\Order(p^0)$.

The results for the individual diagrams of Fig.~\ref{fig:treediagram}
are given in Appendix~\ref{app:tree}. In order to ensure that
we do not miss any effects of the nontrivial analytic structure of
triangle graphs, resulting from the $B^*$ pole terms once
rescattering between the two outgoing pions is taken into account, we keep the full relativistic form of the
denominator part of the propagator.  The latter is connected with the above
heavy-meson approximation Eq.~\eqref{HMpropagator} by~\cite{haiyang}
\begin{align}\label{expansion}
\frac{i}{2v\cdot k}&\longrightarrow \frac{-im_B}{(p_B-k)^2-m_B^2} \,, \nl
\frac{i}{2(v\cdot k+\Delta)} &\longrightarrow \frac{-im_{B^*}}{(p_B-k)^2-m_{B^*}^2} \,, 
\end{align}
where $p_B=m_B v$ is the on-shell $B$ meson momentum.
Written in terms of $s$ and $s_l$, the
pole terms can then be easily identified as
\begin{align}
F^{\textrm{pole}}&=R^{\textrm{pole}}-G^{\textrm{pole}}\,, \quad
R^{\textrm{pole}}=\frac{\alpha}{u-m_{B^*}^2}\,,\nl
F_2^{\textrm{pole}}&=G^{\textrm{pole}}=\frac{\beta}{u-m_{B^*}^2}\,, ~
F_3^{\textrm{pole}}=H^{\textrm{pole}}=\frac{\gamma}{u-m_{B^*}^2}\,,\nl
F_1^{\textrm{pole}}&=X\cdot F_{\textrm{pole}}+\sigma_\pi(PL)\cos\theta_\pi G_{\textrm{pole}}\nl 
&=\frac{X(\alpha-\beta)+\sigma_\pi(PL)\cos\theta_\pi \beta}{u-m_{B^*}^2}\,,
\label{eq:Fipoles}
\end{align}
using the abbreviations 
\begin{align}
\alpha&\equiv -\frac{g^2f_Bm_B^2m_{B^*}}{f_\pi^2(m_B^2-s_l)}\big(s-2M_\pi^2\big)\,,\nl
\beta&\equiv-\frac{gf_Bm_B^2m_{B^*}}{2f_\pi^2}\,, \quad
\gamma\equiv-\frac{g^2f_Bm_B^3m_{B^*}^2}{f_\pi^2(m_{B^*}^2-s_l)}\,.
\label{alphas}
\end{align}
All pole contributions start to contribute at the expected leading chiral orders.
We note, though, that $\alpha=\Order(p)$ is subleading to $\beta=\Order(p^0)$ 
in $F^{\textrm{pole}}$ and $F_1^{\textrm{pole}}$, and can be neglected; they are indeed partially
an artifact of the translation of the heavy-meson formalism back into relativistic kinematics
in the calculation of Ref.~\cite{Dl4}. We will use the contributions
$\propto \alpha$ in the partial waves $f_i$ later on to illustrate potential higher-order effects, 
although these are neither complete nor necessarily dominant amongst the subleading contributions
(cf.\ the discussion of the scaling behavior of higher-order terms in the current in Ref.~\cite{Burdman}).
For the purpose of the ($s$-channel) partial-wave projections to be performed later,
the $u$-channel pole can be written in terms of $s$ and $\cos\theta_\pi$,
\begin{align}
u(s,\cos\theta_\pi)-m_{B^*}^2 & =\sigma_\pi X (\cos\theta_\pi + y)\,,\nl 
y &=\frac{\Sigma_0-s-2m_{B^*}^2}{2\sigma_\pi X} \,.
\end{align}

Finally, the remaining, nonpole, parts of the amplitude can also be extracted from the 
expressions in Appendix~\ref{app:tree}.  There are nonvanishing contributions to the form 
factor $F_1$ only, which in view of the required partial-wave expansion we write as
\begin{align}\label{eq:matchfunctions}
\frac{F_1(s)^{\textrm{$\chi$PT}}-F_1^{\textrm{pole}}}{X} & =M_0(s)^{\textrm{$\chi$PT}}
+\frac{2\sigma_\pi \cos\theta_\pi}{X}M_1(s)^{\textrm{$\chi$PT}}\,,\nl
M_0(s)^{\textrm{$\chi$PT}}&=-\frac{(1-g)^2f_Bm_B}{4f_\pi^2} \,,\nl
M_1(s)^{\textrm{$\chi$PT}}&=\frac{(1-g^2)f_Bm_B}{4f_\pi^2(m_B^2-s_l)}X^2 \,.
\end{align}
$M_0(s)^{\textrm{$\chi$PT}}$ and $M_1(s)^{\textrm{$\chi$PT}}$ are found to be of chiral orders $p^0$ and $p$, respectively,
and therefore suppressed by one order compared to the pole terms~\cite{Burdman},
as explained in Appendix~\ref{app:tree}. 
We will use these expressions in Sec.~\ref{sec:matching} to match
the polynomial parts of the dispersive representations of the corresponding amplitudes,
but again rather in order to illustrate potential uncertainties due to subleading effects:
these contributions are not complete even at the chiral order at which they occur.

To conclude this section, we point out that in order for the
chiral counting scheme to work consistently, we have to assume the
lepton invariant mass squared $s_l$ to be large, of the order of
$m_B^2$. This limits the kinematic range of applicability of our
approach to match the dispersive representation derived in the following 
to heavy-meson chiral perturbation theory.

\subsection{Omn\`es representation}\label{Omnes}

Having fixed the tree-level decay amplitude and in particular the pole terms,
we proceed to analyze the effects of pion--pion rescattering using dispersion relations.
This will give access to the $s$~dependence of the decay form factors (roughly up to 1\,GeV,
as detailed in Sec.~\ref{sec:analytic}) in a model-independent way.
We will resort to the formalism based on Omn\`es representations as introduced in Ref.~\cite{eta3pi}.
For its application to the closely related process of $K_{l4}$ decays, see Refs.~\cite{StofferThesis,StofferProc}.
Note, however, that everything discussed in the following is to be understood at fixed $s_l$:
dispersion theory as applied here does not allow us to improve on the form factor dependence
on the dilepton invariant mass, beyond what the chiral representation in the previous section includes.
We emphasize once more that the dispersive aspect of our analysis is in principle independent
of the matching to heavy-meson chiral perturbation theory: the validity of any theoretical description 
of the different form factors in the soft-pion limit ($s\approx 0$) 
can be extended at least to the whole kinematic region of elastic $\pi\pi$ scattering with this method.

We may write an alternative form of the partial-wave expansion
Eq.~\eqref{PWD} for the pole-term-subtracted amplitudes, neglecting terms beyond $P$~waves,
\begin{align}\label{reconstruction}
\frac{F_1(s,t,u)}{X}&=\frac{F_1^{\textrm{pole}}}{X}+M_0(s)-\frac{(t-u)}{X^2}M_1(s)\,,\nl
F_2(s,t,u)&=F_2^{\textrm{pole}}+U_1(s)\,,\nl
F_3(s,t,u)&=F_3^{\textrm{pole}}+V_1(s)\,.
\end{align}
Here and in the following we suppress the dependence on $s_l$, which is kept fixed. 
The additional factor of $X^2$ in the definition of $M_1$ avoids 
the introduction of kinematic singularities at the zeros of $X$
(in particular at the limit of the physical decay region $s=(m_B-\sqrt{s_l})^2$).
The functions $M_0$, $M_1$, $U_1$, and $V_1$ defined
this way possess right-hand unitarity branch cuts as their only
nontrivial analytic structure and no poles. Since the pole terms
$F_1^{\textrm{pole}}/X$, $F_2^{\textrm{pole}}$, $F_3^{\textrm{pole}}$
are real, one immediately finds
\begin{align}
\Im f_0(s)&=\Im  M_0(s)\,,\quad \Im \Big(\frac{X}{2\sigma_\pi}f_1\Big)= \Im M_1(s)\,,\nl \Im g_1(s)&=\Im
U_1(s)\,,\quad \Im h_1(s)=\Im V_1(s)\,, \label{eq:Im-f-M}
\end{align}
which allows us to write
\begin{align}\label{hat}
f_0(s)&=M_0(s)+\hat M_0(s)\,, ~
f_1(s)=\frac{2\sigma_\pi}{X}\big(M_1(s)+\hat M_1(s)\big)\,,\nl 
g_1(s)&=U_1(s)+\hat U_1(s)\,, ~~\,
h_1(s)=V_1(s)+\hat V_1(s)\,.
\end{align}
The real ``hat functions'' $\hat M_0(s)$, $\hat M_1(s)$,
$\hat U_1(s)$, and $\hat V_1(s)$ are the partial-wave projections of
the pole terms given in Eqs.~\eqref{eq:Fipoles}--\eqref{alphas},
which explicitly read
\begin{align}\label{eq:hatfunction}
\hat M_0(s)&=\frac{\xi\,Q_0(y)+(PL)\beta}{X^2}\,,\quad
\hat M_1(s)=-\frac{3\xi}{2\sigma_\pi X}Q_1(y) \,,\nl
\xi&=\frac{X}{\sigma_\pi}(\alpha-\beta)-(PL)y\beta \,,\nl 
\hat U_1(s)&=\frac{\beta}{\sigma_\pi X}\Big(Q_0(y)-Q_2(y)\Big)\,,\nl 
\hat V_1(s)&=\frac{\gamma}{\sigma_\pi X}\Big(Q_0(y)-Q_2(y)\Big)\,,
\end{align}
where the $Q_l(y)$ are Legendre functions of the second kind,
\begin{equation}
Q_l(y)=\frac{1}{2}\int_{-1}^1\frac{dz}{y-z}P_l(z)\,.
\end{equation}
Explicitly, the first three of these read
\begin{align}
Q_0(y)&=\frac{1}{2}\log\frac{y+1}{y-1}\,, \quad
Q_1(y)=yQ_0(y)-1\,,\nl
Q_2(y)&=\frac{3y^2-1}{2}Q_0(y)-\frac{3}{2}y\,. \label{Ql(y)}
\end{align}
We have projected onto the partial waves of $F_2$ and $F_3$ [whose partial-wave expansions proceed in
derivatives of Legendre polynomials---see Eq.~\eqref{eq:pw-expansion}]
using
\begin{equation}
\int_{-1}^1P'_i(z)\big[P_{j-1}(z)-P_{j+1}(z)\big]dz=2\delta_{ij}\,.
\end{equation}
Note that, in order to show that the partial-wave-projected pole terms above indeed
are real everywhere along the right-hand cut, i.e.\ for all $s\geq 4M_\pi^2$, care has to be 
taken about the correct analytic continuation.  For example, 
$X$, only defined unambiguously in the physical decay region in Eq.~\eqref{definitionX},
is continued according to~\cite{Kambor,Bronzan}
\begin{equation}
X=\left\{\begin{array}{ll} |X|\,, & s\in\big[4M_\pi^2,(m_B-\sqrt{s_l})^2\big] \,,\\[1mm]
i|X| \,,& s\in\big[(m_B-\sqrt{s_l})^2,(m_B+\sqrt{s_l})^2\big] \,,\\[1mm]
-|X| \,,& s\in\big[(m_B+\sqrt{s_l})^2,\infty\big) \, \end{array}\right.
\end{equation}
(where the last range is of no practical relevance for our dispersive integrals).
Furthermore, 
in the range of $(m_B-\sqrt{s_l})^2 < s < (m_B+\sqrt{s_l})^2$, 
the argument $y$ of the Legendre functions of the second kind  becomes purely
imaginary; the lowest one can be expressed as $Q_0(y)=i(\pi/2-\arctan|y|)$.
In particular, no singularities arise at the zeros of $X$, $s=(m_B\pm\sqrt{s_l})^2$.
Physically, the reality of the pole terms is based on the fact that the $B^*$ 
cannot go on its mass shell in any kinematic configuration.

In the elastic regime, the right-hand cut of the partial waves $f_i$
($i=0,\,1$), $g_1$, $h_1$ for $s>4M_\pi^2$ is given by discontinuity
equations relating them to the elastic $\pi\pi$ partial-wave
amplitudes $t_i^i(s)$, $i=0,\,1$,\footnote{We use this somewhat
unusual notation owing to the fact that we only consider $S$ and
$P$~waves, and no isospin $I=2$ is allowed.} according to
\begin{align}
\textrm{disc}\,f_i(s) &= f_i(s+i\epsilon)-f_i(s-i\epsilon)=2i\,\Im f_i(s) \nl
&= 2i\sigma_\pi f_i(s)\big[t_i^i(s)\big]^* = f_i(s)e^{-i\delta_i^i(s)}\sin\delta_i^i(s) \,, \label{eq:discf}
\end{align}
where we have expressed $t_i^i(s)$ in terms of the corresponding
phase shift $\delta_i^i(s)$ in the usual way. Analogous equations
hold for $g_1$ and $h_1$.  Equation~\eqref{eq:discf} implies Watson's
theorem: the phase of the partial wave equals the elastic phase
shift. From Eqs.~\eqref{eq:Im-f-M} and \eqref{hat}, one finds
\begin{equation}\label{inhomogeneity}
\Im M_i(s)
=\left(M_i(s)+\hat M_i(s)\right)e^{-i\delta_i^i(s)}\sin\delta_i^i(s) \,,
\end{equation}
and similarly for $U_1(s)$, $V_1(s)$.

Equation~\eqref{inhomogeneity} demonstrates that the hat functions
constitute inhomogeneities in the discontinuity equations.
The solution is given by~\cite{eta3pi}
\begin{align}\label{Omnesrepresentation}
M_i(s)&=\Omega_i^i(s)\bigg\{P_{n-1}(s)\nl&\quad+\frac{s^n}{\pi}\int_{4M_\pi^2}^\infty\frac{\hat
M_i(s')\sin\delta_i^i(s')ds'}{|\Omega_i^i(s')|(s'-s-i\epsilon)s'^n}\bigg\} \,,
\end{align}
where $P_{n-1}(s)$ is a subtraction polynomial of degree $n-1$, and the Omn\`es function is defined as~\cite{Omnes}
\begin{equation}
\Omega_l^I(s)=\exp
\bigg\{\frac{s}{\pi}\int^\infty_{4M_\pi^2}\frac{\delta_l^I(s')ds'}{s'(s'-s-i\epsilon)}\bigg\}\,.
\end{equation}
The standard Omn\`es solution $P_{n-1}(s)\Omega_i^i(s)$ of the homogeneous discontinuity equation
($\hat M_i=0$), valid for form factors without any left-hand pole or cut structures, is modified by a dispersion
integral over the inhomogeneities $\hat M_i$, which in the present case are given by the partial-wave-projected pole terms.

The minimal order of the subtraction polynomial is dictated by the requirement of the dispersive integral
to converge.
First we note that, if the phase $\delta_l^I(s)$ asymptotically approaches a constant value $c\pi$,
then the corresponding Omn\`es function falls off asymptotically  $\sim s^{-c}$.
We will assume both $\pi\pi$ input phases to approach $\pi$ for large energies,
\begin{equation}\label{asy}
\delta_0^0(s) \longrightarrow \pi \,, \quad \delta_1^1(s)\longrightarrow \pi \,,
\end{equation}
such that $\Omega_0^0(s),\,\Omega_1^1(s) \sim 1/s$ for large $s$.

A more problematic question concerns the behavior of the hat functions for large $s$.
In principle, this is entirely determined by the partial-wave-projected $B^*$ pole terms as given in Eq.~\eqref{eq:hatfunction}.
However, as we have decided to include the relativistic pole graphs, these explicitly contain the scale $m_B$,
and the asymptotic behavior is only reached for $\sqrt{s} \gg m_B$---far too high a scale, given that we realistically know
the pion--pion phase shifts only up to well below 2\,GeV, and that we presently neglect all inelastic contributions, 
which set in above 1\,GeV.  
We can formally remedy this problem by just considering the large-$s$ behavior
of the heavy-meson approximation of the pole terms,\footnote{Remember that we made use of the relativistic pole
terms mainly to ensure the correct analytic properties at \emph{low} energies, i.e.\ in the near-threshold region.} 
in which $m_B$ only features parametrically as a prefactor;  
being aware that corrections to the heavy-meson approximation scale like $\sqrt{s}/m_B$, which is not a very
small quantity in the region of $1\GeV \lesssim \sqrt{s} \lesssim 2\GeV$, say.
In the heavy-meson approximation, i.e., at leading order in an expansion of $1/m_B$,
the inhomogeneities of Eq.~\eqref{eq:hatfunction} behave according to
\begin{align}
\hat M_0(s) &\sim s^{-1/2} \,, & \hat M_1(s) &\sim s^0 \,, \nl 
\hat U_1(s) &\sim s^{-1/2} \,, & \hat V_1(s) &\sim s^{-1/2} \,.
\end{align}
Together with the large-$s$ behavior of the Omn\`es functions, we
conclude that the representation for $M_1(s)$ requires at least two subtractions, 
while for $M_0(s)$, $U_1(s)$, and $V_1(s)$, one subtraction each seems to be sufficient.
\begin{figure}
\includegraphics[width=\linewidth,clip=true]{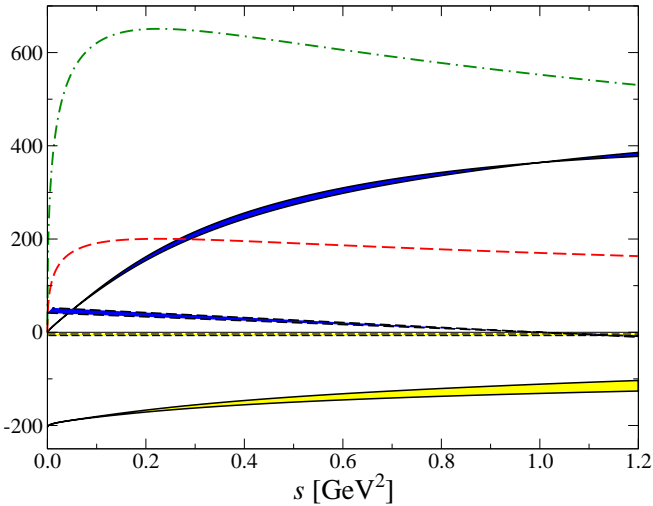}
\caption{Hat functions $\hat M_0(s)$ (yellow band with full lines), $\hat M_1(s)$ (blue band with full lines), 
$\hat U_1(s)$ (red dashed line), and $\hat V_1(s)$ (green dot-dashed line),
for $s_l=(m_B-1\GeV)^2$. 
We also show the polynomial contributions to the form factor $F_1/X$, 
for $S$ (yellow band with dashed lines) and $P$~wave (blue band with dashed lines), 
which are seen to be strongly suppressed.
$\hat M_1(s)$ as well as the $P$-wave polynomial $M_1(s)^{\textrm{$\chi$PT}}$ 
are given in units of $\textrm{GeV}^{-2}$, all other functions are dimensionless. \label{fig:hat}}
\end{figure}
However, looking at the behavior of the various hat functions in the low-energy region
in Fig.~\ref{fig:hat} (for a special value of $s_l = (m_B-1\GeV)^2$), 
we note that the falling of $\hat M_0(s)$, $\hat U_1(s)$, and $\hat V_1(s)$
barely seems to set in in the kinematical region $s \lesssim 1\GeV^2$ 
where we have to assume the spectral function to be saturated, while
$\hat M_1(s)$ even grows at those energies instead of approaching a constant value.
It seems therefore advisable to oversubtract all the dispersive representations once, 
such as to allow for two subtraction constants each for $M_0(s)$, $U_1(s)$, and $V_1(s)$,
and three for $M_1(s)$.  This way, inelastic contributions at higher energies that we do
not take into account explicitly should also be more effectively suppressed.
The complete set of dispersion relations of the Omn\`es type therefore reads
\begin{align}
M_0(s)&=\Omega_0^0(s) \bigg\{a_0+a_1s\nl&\qquad\quad
+\frac{s^2}{\pi} \int_{4M_\pi^2}^\infty \frac{\hat M_0(s')\sin\delta_0^0(s')ds'}{|\Omega_0^0(s')|(s'-s-i\epsilon){s'}^2}\bigg\},\nl
M_1(s)&=\Omega_1^1(s)\bigg\{a'_0+a'_1 s+a'_2s^2\nl&\qquad\quad
+\frac{s^3}{\pi}\int_{4M_\pi^2}^\infty\frac{\hat M_1(s')\sin\delta_1^1(s')ds'}{|\Omega_1^1(s')|(s'-s-i\epsilon){s'}^3}\bigg\},\nl
U_1(s)&=\Omega_1^1(s)\bigg\{b_0 +b_1s\nl&\qquad\quad
+\frac{s^2}{\pi}\int_{4M_\pi^2}^\infty\frac{\hat U_1(s')\sin\delta_1^1(s')ds'}{|\Omega_1^1(s')|(s'-s-i\epsilon){s'}^2}\bigg\},\nl
V_1(s)&=\Omega_1^1(s)\bigg\{c_0+c_1s \nl&\qquad\quad
+\frac{s^2}{\pi}\int_{4M_\pi^2}^\infty\frac{\hat V_1(s')\sin\delta_1^1(s')ds'}{|\Omega_1^1(s')|(s'-s-i\epsilon){s'}^2}\bigg\}.
\label{eq:hatdispersion}
\end{align}
The subtraction constants are \emph{a~priori} unknown, and need to be determined either by further
theoretical input, or by fitting to experimental data. It is easy to
check that the
functions $M_0(s)$, $\ldots$, $V_1(s)$ themselves do not satisfy
Watson's theorem; however, taking into account Eq.~\eqref{hat}, the
partial-wave amplitudes $f_0$ ($f_1$, $g_1$, $h_1$) do; i.e., their
phases equal the elastic scattering phases $\delta_0^0$ ($\delta_1^1$).

We add a few further remarks concerning Fig.~\ref{fig:hat}. 
All of the partial-wave-projected pole terms display singular behavior 
of square-root type at $s=0$ (suppressed as $s^{3/2}$ in the case of $\hat M_1(s)$; 
note that also $\hat M_0(s)$ has a square-root singularity, which is hard to discern 
in Fig.~\ref{fig:hat} due to the axis scaling).
These left-hand singularities obviously carry over to the partial waves: 
close to the $\pi\pi$ threshold, the partial-wave amplitudes cannot be represented by simple scalar or vector
form factors.  

The uncertainty bands for $\hat M_i(s)$, $i=0,\,1$, in Fig.~\ref{fig:hat} indicate the effect of the 
(incomplete) higher-order contribution $\propto \alpha$ in Eq.~\eqref{eq:hatfunction}, suppressed by $1/m_B$ 
and found to be surprisingly small.  We do \emph{not} include the uncertainty due to 
the overall scaling with the coupling constant $g$, which translates directly into an uncertainty
of a projected extraction of $|V_{ub}|$, but does not (at this order) 
affect the shape of the distributions.  The inhomogeneities scale with $g$ according to
$\hat M_i(s),\, \hat U_1(s)\propto g$, $\hat V_1(s) \propto g^2$.

The dispersive method using inhomogeneities as described above has by now been used
for a variety of low-energy processes, such as
$\eta\to3\pi$~\cite{eta3pi,Lanz2},
$\omega/\phi\to 3\pi$~\cite{Franz},
$K\to\pi\pi$~\cite{Orellana},
$K_{l4}$~\cite{StofferThesis,StofferProc},
$\gamma\gamma\to\pi\pi$~\cite{Moussallam,HoferichterPhillips}, or
$\gamma\pi\to\pi\pi$~\cite{Truong,Hoferichter}.
In several of those cases, the inhomogeneities (given in terms of hat functions),
which incorporate left-hand-cut structures, and the amplitudes given in terms of Omn\`es-type solutions
with a right-hand cut only are calculated iteratively from each other, until convergence is reached.
In our present analysis, the ansatz is comparably simpler, as the left-hand cut is approximated by
pole terms, whose partial-wave projections then determine the inhomogeneities.
This is closely related to the method of Ref.~\cite{Moussallam} for $\gamma\gamma\to\pi\pi$,
where the left-hand structures  are approximated by Born terms
and resonance contributions to $\gamma\pi\to\gamma\pi$.

\subsection{Matching the subtraction constants}\label{sec:matching}

We need to consider two essentially different contributions to the subtraction constants
in the representation Eq.~\eqref{eq:hatdispersion}, writing them formally as
\begin{equation}
a_i = \bar a_i + \hat a_i \,,
\end{equation}
and similar decompositions for the $a'_i$, $b_i$, and $c_i$.
We discuss the contributions $\hat a_i$ etc.\ first.
We argue in Appendix~\ref{app:poly-inhom}
that for inhomogeneities of essentially constant ($\hat M_0$, $\hat U_1$, $\hat V_1$)
or approximately linear ($\hat M_1$) behavior over a large part of 
the kinematical region of interest, the coefficients of the highest power in the 
subtraction polynomials ($a_1$, $a'_2$, $b_1$, and $c_1$) need to be adjusted
in order to provide a reasonable high-energy behavior.\footnote{This can be corroborated
to some extent by arguments from Brodsky--Lepage quark counting rules~\cite{BrodskyLepage} 
and soft-collinear effective theory~\cite{SCET}, albeit in kinematic regions 
with completely different scaling of $s_l$ with respect to $m_B^2$ (taken as fixed and not particularly large here). 
Assuming the large-$s$ behavior of the different form factors and partial waves is independent thereof, 
we indeed need to require the leading powers in $s$ to cancel between the dispersion integrals over the inhomogeneities
and the subtraction polynomial.}  These coefficients are given
by the derivative of the corresponding Omn\`es function at $s=0$, multiplied with 
the constant/the derivative of the inhomogeneity in question.  Obviously, the hat functions
are not exactly constant/linear: to the contrary, they include square-root singularities at
$s=0$ due to the left-hand cut.  There is, therefore, necessarily an uncertainty due to the
choice of a ``matching point'' 
$s_m$ at which to evaluate these ``constants,''  
\begin{align}
\hat a_1 &= \hat M_0(s_m) \times \dot\Omega_0^0(0) \,, &
\hat a_2' &= \frac{\hat M_1(s_m)}{s_m} \times \dot\Omega_1^1(0) \,, \nl
\hat b_1 &= \hat U_1(s_m) \times \dot\Omega_1^1(0) \,, &
\hat c_1 &= \hat V_1(s_m) \times \dot\Omega_1^1(0) \,. \label{eq:abc-Omnesdot}
\end{align}
We choose $s_m=M_\rho^2$, due to the expected strong enhancement of the distribution at the $\rho$ resonance peak.
Here, $\dot\Omega_l^I(0) =d\Omega_l^I(s)/ds|_{s=0}$.  All other subtraction constants do not
receive ``hat'' contributions.

The second contribution to the subtraction constants, dominantly to those of \emph{low}
polynomial order in $s$, stems from matching to the nonpole part of the chiral amplitude Eq.~\eqref{eq:matchfunctions}, 
which yields (for fixed $s_l$) a polynomial contribution in $s$. 
In this exploratory study we use the leading-order expressions only. We
expect the chiral expansion to converge best at the sub-threshold
point $s=0$, as opposed to, e.g., the $\pi\pi$
threshold~\cite{Colangelo}.

As we match the dispersive representation Eq.~\eqref{eq:hatdispersion}
to the leading chiral tree-level amplitude, which does not
contain any rescattering or loop corrections, we identify the
subtraction constants $\bar a_{0-1}$, $\bar a'_{0-2}$ by setting the scattering phases to
zero, i.e., $\Omega_i^i(s)\equiv 1$, and the dispersive
integrals over the inhomogeneities vanish. At $s=0$, we find from Eq.~\eqref{eq:matchfunctions}
\begin{align}
\bar a_0 &=-\frac{(1-g)^2f_Bm_B}{4f_\pi^2} \,, &
\bar a_1 &= 0\,,\nl
\bar a'_0 &= \frac{(1-g^2)f_Bm_B}{16f_\pi^2}\big(m_B^2-s_l\big)\,, && \nl 
\bar a'_1 &= -\frac{(1-g^2)f_Bm_B}{8f_\pi^2}\frac{m_B^2+s_l}{m_B^2-s_l}\,, &
\bar a'_2 &= \frac{(1-g^2)f_Bm_B}{16f_\pi^2(m_B^2-s_l)}\,. \label{eq:bar-a}
\end{align}
The term $\propto \bar a'_2 s^2$, stemming from the expansion of $X^2$, is chirally suppressed
and could as well be neglected.
$F_2$ and $F_3$ at leading order coincide with their pole terms,
thus the matching implies that the parameters $\bar b_i$ and $\bar c_i$ vanish.

In order to illustrate the relative importance of the (partial-wave-projected) pole terms
relative to the subtraction polynomial---that is, the decompositions $M_i(s)+\hat M_i(s)$
on tree level, for $i=0,\,1$---we also show these for $s_l=(m_B-1\GeV)^2$ in 
Fig.~\ref{fig:hat}.
We verify the expected dominance of the pole terms/the hat functions in $f_0(s)$ and $f_1(s)$, 
as suggested by power counting arguments.  For the uncertainty bands
of the polynomial corrections with mixed dependence on $g$,
we have varied this coupling within its assumed uncertainty, $g=0.58\pm0.07$.

Remember that $g_1(s)$ and $h_1(s)$ consist of $B^*$ pole terms only at leading order: 
this pole dominance should have very favorable consequences for the 
reliability of the form factor prediction, as the pole contributions are essentially fixed by the coupling constant $g$ 
(as well as $f_B$) beyond the chiral expansion; the latter affects only the precision
of the polynomial contribution.
Next-to-leading-order corrections to the residues of the pole terms seem to have surprisingly little effect.

\section{Results}\label{sec:results}

\subsection{Scattering phase input}\label{sec:phases}

The $\pi\pi$ phase shifts are known to sufficient accuracy in the
region $s\lesssim s_0 \equiv(1.4\, \GeV)^2$ (cf.\ Refs.~\cite{ACGL,Pelaez}).
In order to ensure the assumed asymptotic behavior $\delta_0^0(s),\,\delta_1^1(s) \to \pi$
for $s\to\infty$, we continue the phases beyond $s_0$ according to the prescription~\cite{MoussallamNf}
\begin{equation}\label{extrapolation}
\delta_i^i(s\geq s_0)=\pi + \big(\delta_i^i(s_0)-\pi\big)f\Big(\frac{s}{s_0}\Big) \,, ~
f(x) = \frac{2}{1+x^{3/2}} \,.
\end{equation}
There is a further subtlety concerning the $S$-wave phase shift: as
we have discussed in Sec.~\ref{sec:analytic}, the elastic
approximation breaks down at the $K\bar K$ threshold $s_K$ with the
occurrence of the $f_0(980)$ resonance. Both the phase of the
partial wave $\arg t_0^0(s)$ and, e.g., the phase of the
nonstrange scalar form factor of the pion $\arg F_\pi^S(s)$ differ
significantly from $\delta_0^0(s)$ in this region: they quickly drop
and then roughly follow the energy dependence of $\delta_0^0(s)$ again, with
$\delta_0^0(s) - \arg t_0^0(s) \approx \delta_0^0(s) - \arg
F_\pi^S(s) \approx \pi$~\cite{Anant}. Therefore a single-channel approximation to the pion scalar form factor
only works for $s<s_K$ if a phase of the form of either $\arg
t_0^0(s)$ or $\arg F_\pi^S(s)$ are used as input to the Omn\`es
function instead of $\delta_0^0(s)$. We use such a form factor phase
taken from Ref.~\cite{Ditsche}. Obviously, we cannot provide a
reliable description of pion--pion rescattering effects where the
inherent two-channel nature of the problem becomes important, hence
our dispersive description is confined to below $s_K$.

With the phase shift input thus continued formally up to infinity,
the Omn\`es integrals can be fully performed. 
We have checked that different continuation prescriptions from
the one given in Eq.~\eqref{extrapolation} above $s_0$
have very little impact on the physics at low energies, i.e., below 1\,GeV.

The phase input allows us to evaluate the derivatives of the Omn\`es functions
required in Eq.~\eqref{eq:abc-Omnesdot} via the sum rules
\begin{equation}
\dot\Omega_l^I(0) = \frac{1}{\pi}\int_{4M_\pi^2}^\infty ds' \frac{\delta_l^I(s')}{{s'}^2} \,,
\end{equation}
leading to $\dot\Omega_0^0(0) = 2.5\GeV^{-2}$, $\dot\Omega_1^1(0) = 1.8\GeV^{-2}$.
This corresponds to squared radii of the pion scalar and vector form factors
$\langle r_S^2\rangle = 0.58\,\text{fm}^2$, 
$\langle r_V^2\rangle = 0.42\,\text{fm}^2$, 
both only around 5\% below the central values of more sophisticated evaluations~\cite{CGL,BijTal,HanhartFF}.

In order to ensure numerically stable results, we perform the
dispersion integrals over the inhomogeneities
Eq.~\eqref{eq:hatdispersion} up to $\sqrt{s} = 3\,$GeV. This upper
limit of the integration does not have any real physical
significance: it merely represents an attempt to sum up the
high-energy remainder of the integral to reasonable approximation,
and does not mean we pretend to understand $\pi\pi$ interactions at
such scales.

\subsection{Subtraction constants, spectrum}\label{sec:Nsubtraction}

We illustrate the results of our discussion for a sample value of $s_l=(m_B-1\GeV)^2$,
which means the kinematically allowed range in the invariant mass of the pion pair
extends to $\sqrt{s} = 1\GeV$. 
Evaluating the (nonvanishing) subtraction constants obtained from matching to the nonpole, polynomial
parts of the chiral tree-level amplitude, Eq.~\eqref{eq:bar-a}, we find
\begin{align}
\bar a_0&=-5.3\pm1.8\,, &  
\bar a'_0&=(48\pm6)\GeV^2\,, \nl 
\bar a'_1&=-48\pm6 \,,  
& \bar a'_2 &=(0.5\pm0.1)\GeV^{-2}\,,
\end{align}
where the errors refer to the uncertainty in $g$ only.
The ``hat'' contributions to the subtractions of Eq.~\eqref{eq:abc-Omnesdot}, at $s_l=(m_B-1\GeV)^2$, 
are found to be
\begin{align}
\hat a_1 &=  (-363\ldots-330) \Big(\frac{g}{0.58}\Big)\GeV^{-2}\,, \nl
\hat a_2'&=  (888\ldots924)\Big(\frac{g}{0.58}\Big)\GeV^{-2}\,, \nl
\hat b_1 &=  332\Big(\frac{g}{0.58}\Big)\GeV^{-2}\,, \quad
\hat c_1 =  1078\Big(\frac{g}{0.58}\Big)^2\GeV^{-2}\,,
\end{align}
where we have displayed the scaling with $g$ explicitly and shown the range of parameters
in the $F_1$ partial waves due to the higher-order corrections discussed above.

\begin{figure}
\centering
\includegraphics[width=\linewidth]{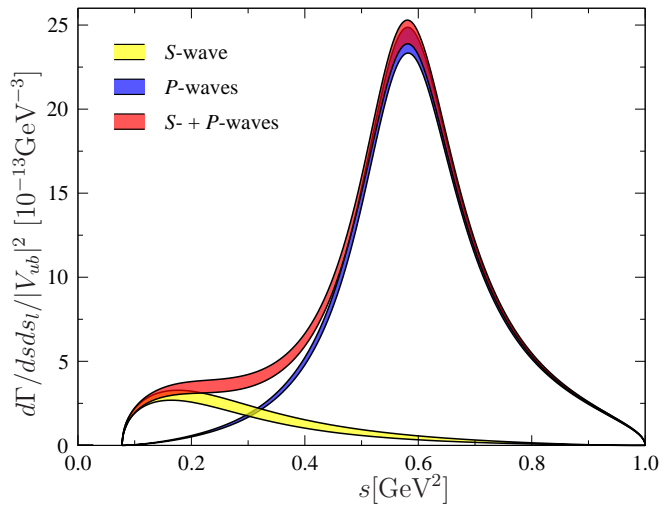}
\caption{Differential decay width $d\Gamma/ds ds_l$ divided by $|V_{ub}|^2$ for
the example value of $s_l=(m_B-1\GeV)^2$, 
decomposed into $S$- and $P$-wave contributions.  For details, see discussion in main text.}
\label{fig:demonstration}
\end{figure}

For demonstration, we plot the partial decay rate in Fig.~\ref{fig:demonstration}
for the dilepton invariant mass $s_l = (m_B-1\GeV)^2$.
We find that the $S$-wave contribution leads to a significant enhancement of the
spectrum at low $\pi\pi$ invariant masses, beyond what might be considered $\rho$ dominance.
The near-threshold dominance of the $S$~wave was already pointed out 
in Ref.~\cite{Burdman} in the context of heavy-meson chiral perturbation theory.
Concerning the different $P$~waves, we find that the kinematical prefactor $X^2/m_B^4$ strongly 
suppresses the partial wave $h_1$ or the form factor $F_3$ for the values of $s_l$ considered here.
Of the other two, $g_1$ yields a contribution to the differential rate roughly twice as large as $f_1$.

\section{Discussion and summary}\label{sec:summary}

We wish to emphasize that matching to chiral perturbation theory at
leading order can only be considered an estimate, and mainly
serves for illustration purposes here. Higher-order corrections are
expected to be significant. 
Ultimately, the subtraction constants that influence the shape ought to be determined by fits to
experimental data; they can be thought of as parametrizing a ``background polynomial,''
beyond the dominant pole terms, albeit with completely correct rescattering corrections, 
obeying Watson's theorem.
The necessary theoretical normalization of the form factors is essentially provided 
at $s = M_\rho^2$, via Eq.~\eqref{eq:abc-Omnesdot}; its stability under higher-order corrections
still merits further investigation in order to provide a theoretical uncertainty for
$|V_{ub}|$ extracted from $B_{l4}$ decays.

To summarize, we have provided a description of the form factors for
the decay $B^-\to \pi^+\pi^- l^-\bar\nu_l$ using dispersion theory,
which should lead to an improved method to measure $|V_{ub}|$.
Pion--pion final-state interactions have been included
nonperturbatively in the elastic approximation, while left-hand-cut
structures in the $\pi B$ interaction are approximated by $B^*$ pole
terms. We stress that our formalism allows, for the first time, to
use the full information for $\pi\pi$ invariant masses below 1\,GeV,
without the need to refer to particular parametrization for selected
resonances such as the $\rho(770)$ [or the $f_0(980)$]; 
it allows for a full exhaustion of the corresponding spectra.
Improved experimental data to allow for such an analysis to be performed in
practice is therefore highly desirable.

As an outlook concerning theoretical improvement, we have hinted at the possibility 
to extend the present analysis to lower values of the dilepton invariant mass $s_l$, 
beyond the range of applicability of heavy-meson chiral perturbation theory, but still
making use of dispersion relation for the dependence on the dipion invariant mass $s$.
One promising constraint could be obtained from soft-pion theorems~\cite{softpion}, 
which relate linear combinations of $B_{l4}$ form factors at $s=M_\pi^2$,
but arbitrary $s_l$, to $B\to\pi l\nu$ ($B_{l3}$) form factors at same $s_l$.  Given reliable phenomenological
information on the form factors for $B_{l3}$, this may provide precisely (part of) the matching information
needed to extend the dispersive method of this paper to lower values of $s_l$.  

\begin{acknowledgments}
We would like to thank Thomas Mannel for introducing us to the importance of this problem,
and Johanna Daub, S\'ebastien Descotes-Genon, Jochen Dingfelder, 
Martin Hoferichter, and Peter Stoffer for useful discussions and helpful comments on the manuscript.
This work is supported in part by the DFG and the NSFC through
funds provided to the Sino-German CRC 110 ``Symmetries and the Emergence of Structure in QCD.''
\end{acknowledgments}

\appendix

\section{Tree-level amplitudes in heavy-meson chiral perturbation theory}
\label{app:tree}

Calculating the tree-level diagrams in Fig.~\ref{fig:treediagram}
in heavy-meson chiral perturbation theory, 
one obtains the corresponding amplitudes~\cite{Dl4} [$\cal{A}$--$\cal{D}$, in obvious correspondence to diagrams (a)--(d)]
\begin{align}
\cal{A}&=\frac{if_B}{4f_\pi^2}p_B^\mu\,,\qquad \cal{B}=ip_-^\mu
\cal{B}^{(1)}+ip_B^\mu \cal{B}^{(2)}\,,\nl
\cal{B}^{(2)}&=-\frac{gf_B}{2f_\pi^2}\frac{v\cdot p_-}{v\cdot p_- + \Delta} 
= - \frac{v\cdot p_-}{m_B} \cal{B}^{(1)} \,,\nl 
\cal{C}&= ip_B^\mu\cal{C}^{(1)}+\epsilon^{\mu\alpha\beta\gamma}p_{B\alpha}p_{-\beta}p_{+\gamma}\cal{C}^{(2)}\,,\nl
\cal{C}^{(1)}&= -\frac{g^2f_B}{2f_\pi^2}\frac{p_+\cdot p_--(v\cdot p_+)(v\cdot p_-)}{[v\cdot(p_++p_-)][v\cdot p_- + \Delta]}\,,\nl 
\cal{C}^{(2)}&= -\frac{g^2f_B}{2f_\pi^2}\frac{1}{[v\cdot(p_++p_-)+\Delta][v\cdot p_-+ \Delta]}\,,\nl 
\cal{D}&=  ip_B^\mu\cal{D}^{(1)}\,,\quad
\cal{D}^{(1)}=-\frac{f_B}{4f_\pi^2}\frac{v\cdot(p_+-p_-)}{v\cdot(p_++p_-)}\,.
\end{align}
Identifying the contributions to the individual decay form factors,
we find for these as the leading-order (LO) results 
\begin{align}
F^{\textrm{LO}}&=R^{\textrm{LO}}\!-G^{\textrm{LO}},~
G^{\textrm{LO}}=\frac{m_B}{2}\cal{B}^{(1)},~
H^{\textrm{LO}}=-\frac{m_B^3}{2}\cal{C}^{(2)} ,\nl
R^{\textrm{LO}}&=-\frac{m_Bf_B}{4f_\pi^2}-m_B\Big(\cal{B}^{(2)}+\cal{C}^{(1)}+\cal{D}^{(1)}\Big)\,.
\end{align}
From these, it is then straightforward to identify the pole
contributions given in Eq.~\eqref{eq:Fipoles}, as well as the nonpole pieces of Eq.~\eqref{eq:matchfunctions}.

It is obvious that all diagrams (a)--(d) are formally of $\Order(p^0)$ in terms of soft pion momenta.
Note, however, that all pieces proportional to $p_B^\mu = P^\mu+L^\mu$ are effectively suppressed:
the part $\propto L^\mu$ enters the form factor $R$, which is suppressed by the small lepton mass 
and neglected throughout the main text, while the part $\propto P^\mu$ leads to a chiral suppression by one order
(and is at least partially an artifact of the heavy-meson approximation anyway).
As a consequence, the only leading contributions are given by the amplitudes $\cal{B}^{(1)}$ and $\cal{C}^{(2)}$
in the above, and hence the $B^*$ pole graphs.  This was already pointed out in Ref.~\cite{Burdman}.

\section{Dispersive representations for polynomial inhomogeneities}
\label{app:poly-inhom}

Consider a partial wave $f(s)$ given at tree level as a constant,
$f^{\rm tree}(s) = A$.  In this case, we can write down the dispersive representation
including final-state interactions right away, \emph{if} we assume a certain high-energy
behavior of the amplitude, as
\begin{equation}
f(s) = A\,\Omega(s) \,, \label{eq:a*Omnes}
\end{equation}
with the Omn\`es function $\Omega(s)$. Here, we assume (as in the main text) an Omn\`es function
falling according to $1/s$, i.e.\ given by a phase shift approaching $\pi$ asymptotically, and a partial
wave that vanishes in the same way for large $s$.  This assumption prevents us from multiplying 
$\Omega(s)$ with a polynomial of higher degree.
  
However, in the spirit of the solution discussed in the main text, it should also be possible
to treat this constant as an inhomogeneity and reconstruct the same solution from the
corresponding formalism.  Our solution is then of the form
\begin{equation}
f(s) = A + \Omega(s) \bigg\{ a + a's + \frac{s^2}{\pi}\int_{4M_\pi^2}^\infty 
\frac{A \sin\delta(s')ds'}{|\Omega(s')| {s'}^2(s'-s)} \bigg\}  \,, \label{eq:f}
\end{equation}
where we have chosen the minimal number of subtractions (two) required to make the 
dispersion integral converge.  Note that the subtraction constants $a$, $a'$ are not 
\emph{a~priori} fixed from the tree-level input;  
we can set $a=0$ by requiring the normalization of the amplitude at $s=0$ to match 
the tree-level input.
The integral in Eq.~\eqref{eq:f}
can be performed explicitly, using a dispersive representation of the 
inverse of the Omn\`es function
\begin{equation}
\Omega^{-1}(s) = 1 - \dot\Omega(0)\,s - \frac{s^2}{\pi} \int_{4M_\pi^2}^\infty \frac{\sin\delta(s')ds'}{|\Omega(s')| {s'}^2 (s'-s)} \,,
\end{equation}
where 
$
\dot\Omega(0) = {d\Omega(s)}/{ds}|_{s=0} 
$.
As a result, we find
\begin{equation}
f(s) = \Omega(s) \Big\{A + \big[a'- A \,\dot\Omega(0)\big]s \Big\} \,.
\end{equation}
Therefore, Eq.~\eqref{eq:a*Omnes} is reproduced if we choose $a=0$, $a'=A\,\dot\Omega(0)$.  
We essentially apply the same requirement on the high-energy behavior as in Eq.~\eqref{eq:a*Omnes}:
terms that do not vanish for large $s$ are only canceled for this specific choice of $a'$.

More generally, if we match to a tree-level amplitude of the form
$A\,s^n$, demanding the same leading behavior near $s=0$ such that all subtraction terms
$\propto s^{m\leq n}$ can be put to zero,
the solution using this tree-level input as an inhomogeneity,
\begin{equation}
A\,s^n + \Omega(s) \bigg\{ a's^{n+1}  + \frac{s^{n+2}}{\pi}\int_{4M_\pi^2}^\infty 
\frac{A\,{s'}^n \sin\delta(s')ds'}{|\Omega(s')| {s'}^{n+2}(s'-s)} \bigg\}  , 
\end{equation}
agrees with the ``canonical'' solution $A\,s^n\,\Omega(s)$,
with the ``correct'' high-energy behavior, only if $a'=A\,\dot\Omega(0)$.

\end{document}